\documentstyle[aps,prl,epsfig,preprint]{revtex}
\begin{document}

\title{Metastability, negative specific heat and weak mixing in 
classical long-range many-rotator system}
\author{Benedito J.C. Cabral$^{a,b}$ and Constantino Tsallis$^{b,c}$
\thanks{ben@adonis.cii.fc.ul.pt, tsallis@cbpf.br} \\
$^a$ Departamento de Qu\'{\i}mica e Bioqu\'{\i}mica, Faculdade
de Ci\^encias da Universidade de Lisboa, Edif\'{\i}cio C8, 1749-016,
Lisboa, Portugal \\
$^b$Centro de F\'{\i}sica da Mat\'eria Condensada, Universidade de Lisboa \\
 Av. Prof. Gama Pinto 2, 1649-003 Lisboa, Portugal \\
$^c$Centro Brasileiro de Pesquisas F\'{\i}sicas, 
Rua Xavier Sigaud 150 \\
22290-180 Rio de Janeiro-RJ, Brazil}

\maketitle

\begin{abstract}

We perform a molecular dynamical study of the isolated $d=1$ classical 
Hamiltonian 
${\cal H} = \frac{1}{2} \sum_{i=1}^N L_i^2 + 
\sum_{i \ne j} \frac{1-cos(\theta_i-\theta_j)}{r_{ij}^\alpha}\
;(\alpha \ge 0)$, known to exhibit a second order phase transition, 
being disordered for $u \equiv U/N{\tilde N} \ge u_c(\alpha,d)$ and 
ordered otherwise 
($U\equiv$ total energy and 
${\tilde N} \equiv \frac{N^{1-\alpha/d}-\alpha/d}{1-\alpha/d}$).
We focus on the nonextensive case $\alpha/d \le 1$ and observe that, 
for $u<u_c$, a basin of attraction exists for the initial conditions
for which the system quickly relaxes onto a longstanding metastable 
state (whose duration presumably diverges with $N$ like ${\tilde N}$)
which eventually crosses over to the  microcanonical Boltzmann-Gibbs
stable state. The temperature associated with the (scaled) average 
kinetic energy per particle is lower in the metastable state than 
in the stable one. 
It is exhibited for the first time that the appropriately scaled 
maximal Lyapunov exponent
$\lambda_{u<u_c}^{max}(metastable) \propto N^{-\kappa_{metastable}} \
;(N \to \infty)$, where, for all values of $\alpha/d$, 
$\kappa_{metastable}$ numerically coincides with {\it one third} of
its value for $u>u_c$, hence decreases from 1/9 to zero when
$\alpha/d$ increases from zero to unity, remaining zero thereafter. 
This new and simple {\it connection between anomalies above and below 
the critical point} reinforces the nonextensive universality scenario.

\end{abstract}

\bigskip

The foundations of statistical mechanics, hence of thermodynamics, 
is a subtle and fascinating matter which has driven enriching 
controversies and clarifications since more than one century
(see, for instance, Einstein's remark on the Boltzmann principle
\cite{einstein}). 
The field remains open to new aspects and proposals.
One of these is nonextensive statistical mechanics, 
proposed in 1988 \cite{tsallis} (see \cite{tsallisreview} for reviews).
This formalism is based on an entropic index $q$ (which recovers 
usual statistical mechanics for $q=1$), and has been applied to a 
variety of systems, covering certain classes of both (meta)equilibrium
and nonequilibrium phenomena, e.g., turbulence \cite{turbulence}, 
hadronic jets produced by electron-positron annihilation \cite{bediaga},
cosmic rays \cite{cosmic}, motion of {\it Hydra viridissima} \cite{arpita}, 
among others. In addition to this, it has been advanced that
it could be appropriate for handling some aspects of long-range
interacting Hamiltonian systems. 
This possibility is gaining plausibility nowadays, 
as argued in \cite{latorarapisardatsallis} and elsewhere. 
Indeed, in molecular dynamical approaches  of isolated systems, strongly
nonmaxwellian velocity distributions have recently been observed
that are consistent with such possibility \cite{latorarapisardatsallis}.
A paradigmatic system in the realm of this discussion is the following
classical Hamiltonian:
\begin{equation}
{\cal H} = \frac{1}{2} \sum_{i=1}^N L_i^2 +
 \sum_{i \ne j} \frac{1-cos(\theta_i-\theta_j)}{r_{ij}^\alpha}\;\;\;\;
 (\alpha \ge 0).
\end{equation}
The inertial planar rotators (ferromagnetic $XY$-like model)
are localized at the sites of a $d$-dimensional periodic lattice.
As distance $r_{ij}$ (measured in crystal units) for a given pair $(i,j)$
we consider the shortest among all the possible ones (due to periodicity).
For $d=1$ we have $r_{ij}=1,2,3,...$; for $d=2$ we have $r_{ij}=1,
\sqrt{2}, 2,...$; for $d=3$ we have $r_{ij}=1,\sqrt{2}, \sqrt{3},2,...$,
and so on. 
The so called $HMF$ system \cite{HMF} is recovered for $\alpha/d=0$,
and the first-neighbor model is recovered for $\alpha/d \to \infty$. 
Hamiltonian (1) is extensive if $\alpha/d>1$ and nonextensive
if $0 \le \alpha/d \le1$. This can be seen as follows. If we define
\begin{equation}
{\tilde N} \equiv 1+d \int_1^{N^{1/d}} dr\; r^{d-1}r^{-\alpha}
 = \frac{N^{1-\alpha/d}-\alpha/d}{1-\alpha/d},
\end{equation}
it can be easily checked that the energy scales as $N{\tilde N}$, i.e.,
it is asymptotically proportional to $N$ if $\alpha/d>1$,
to $N \ln N$ if $\alpha/d=1$, and to $N^{2-\alpha/d}$ if $0 \le \alpha/d<1$.
This Hamiltonian is sometimes presented in the literature in the 
following form:
\begin{equation}
{\cal H^\prime} = \frac{1}{2} \sum_{i=1}^N L_i^{\prime 2} + \frac{1}{{\tilde N}}\sum_{i \ne j} \frac{1-cos(\theta_i-\theta_j)}{r_{ij}^\alpha}\;\;\;\;(\alpha \ge 0).
\end{equation}
which artificially makes its energy to scale as $N$, $\forall (\alpha/d)$.
The transformation from this form to the one presented in Eq. (1),
adopted from now on in the present work,
has been described in detail in \cite{anteneodo}. 
This system has since long been shown \cite{fisher} to obey 
Boltzmann-Gibbs (BG) statistical mechanics for $\alpha/d >1$.
What happens for $0 \le \alpha/d \le 1$ is a subtle question which
is under intensive study nowadays
\cite{tamaritanteneodo,giansantiPRE,giansanti,firporuffo,celiaraul,giansanticagliari}. 
In fact, several long-range-interacting systems are since long known
\cite{thirring,lynden-bell,posch,TCU,sota}
to present a variety of thermodynamical anomalies,
such as negative specific heat and superdiffusion among others.
The molecular dynamics in the isolated Hamiltonian (1) with total 
energy $U$ exhibits, for infinitely large time,
the existence of a second order phase transition at
$u \equiv U/N{\tilde N} = u_c(\alpha,d)$.
For $u \ge u_c$ the system is disordered (paramagnetic-like);
otherwise, it is ordered (ferromagnetic-like).
It exhibits anomalies on both sides of the critical point.

For $u>u_c$ (i.e., in the disordered phase), after a quick transient,
the one-particle distribution of velocities gradually becomes
Maxwellian in the $N \to \infty$ limit, in accordance to what
is expected within BG statistical mechanics.
{\it However}, while $N$ increases, the entire Lyapunov spectrum
approaches zero, which is a quite anomalous behavior;
indeed, no such weakening of chaos is expected nor observed
for $\alpha/d >1$.
This weakening of the sensitivity can be characterized through
the maximal Lyapunov exponent $\lambda_{u>u_c}^{max}$
(appropriately scaled as indicated in \cite{anteneodo}) which,
in the $N \to \infty$ limit, vanishes as
$\lambda_{u>u_c}^{max} \propto N^{-\kappa_d}$ ($d$ stands 
for disordered phase); 
$\kappa_d$ decreases from1/3 to zero while $\alpha/d$ 
increases from zero to unity, and remains zero thereafter \cite{giansanti}.
The fact that the sensitivity to the initial conditions 
becomes sub-exponential (possibly a power-law)
strongly reminds what has been observed
\cite{mapetc1,mapetc2,mapetc3,mapetc4,mapetc5,mapetc6,mapetc7,mapetc8,mapetc9}
in a variety of low dimensional maps,
which are known to be adequately described within
nonextensive statistical mechanical concepts.

For $u<u_c$, after a quick transient, 
the behavior depends from the initial conditions.
Two wide basins of attraction exist in the space of the initial conditions.
One of them (which includes Maxwellian velocity distribution and
all angles equal) yields a standard BG microcanonical distribution
which approaches the BG canonical one in the limit $N\to\infty$.
The other one (which includes waterbag and double waterbag 
velocity distribution and all angles equal) yields
a longstanding metastable (quasi-stationary) state 
(whose associated magnetization is basically zero) and only
at very large time joins the BG distribution
(whose associated magnetization is nonzero).
The duration $\tau$ of this metaequilibrium state diverges with $N$.
It has been conjectured \cite{tsalliscatania} that it does so 
as $\tau \propto {\tilde N}$.
Recent results support this scaling; indeed, (i) for $\alpha = 0$,
this conjecture implies $\tau \propto N$,
which has been verified \cite{latorarapisardatsallis},
(ii) for fixed $N$, it implies that $\tau$ exponentially decays 
with $\alpha$, which once again has been verified \cite{giansanticagliari}.

Our focus in this paper is on the metastable state of the $d=1$ model,
which we study for typical values of $(\alpha,u,N)$.
The time evolution of the model has been generated integrating the
equations of motion through a 4$^{\rm th}$ order symplectic algorithm 
\cite{Yoshida} with a relative error in the total energy conservation
less than 10$^{-4}$.
We verify that the time evolution of the (scaled) average
kinetic energy per particle (which plays the role of temperature)
exhibits two plateaux, the first one being anomalous
and the second one being of the BG class.
This is illustrated in Fig. 1 for $\alpha=0.6$.
In the same figure we show the time evolution of $\lambda_{u<u_c}^{max}$.
As in the $\alpha=0$ case, one expects also for $0<\alpha/d<1$ 
two plateaux in $\lambda_{u<u_c}^{max}(t)$.
We can see, however, that for $\alpha=0.6$ the difference 
is almost unperceptively small;
it might happen that this difference quickly decreases with $\alpha$,
as it is the case for $\tau$, but such study is out of the scope of
the present work.
The systematic detection of both plateaux in the temperature enabled
the calculation of the caloric curves, as illustrated in Fig. 2.
We clearly see the existence of negative specific heat for the metastable
state, just below $u_c$.
Then by focusing on small time (after the transient nevertheless),
it was possible to calculate the
$N$-dependence of $\lambda_{u<u_c}^{max}(metastable)$
for typical values of $\alpha$. The results are shown in Fig. 3.
We verify that $\lambda_{u<u_c}^{max}(metastable)
\propto N^{-\kappa_{metastable}}$, where $\kappa_{metastable}$
decreases from 1/9 (thus confirming \cite{latorarapisardacagliari})
to zero, while $\alpha$ increases from zero to unity, 
and remains zero thereafter. 
Furthermore, we numerically verify a remarkable property, namely (see Fig. 4)
\begin{equation}
\kappa_{metastable}=\frac{\kappa_d}{3}\;\;\;(\forall \alpha).
\end{equation}
This constitutes the first connection found for this type 
of models between the anomalies {\it below} and {\it above} the critical point. This is a conceptually important point. Indeed, if nonextensive statistical mechanics is relevant for such long-range interacting systems as the velocity distributions presented in \cite{latorarapisardatsallis} seem to suggest, one would expect the model to be somehow associated with a {\it single} value of the entropic index $q$ for all energies, both below and above possible critical points. Eq. (4) makes this possibility plausible. Before ending let us mention that no anomalies were detected nor expected for $\lambda_{u<u_c}^{max}(stable)$ (i.e., in the BG regime emerging at large time), which should gradually become positive $N$-independent values for all values of $\alpha$. This is of course consistent with the picture that $\lim_{t \to\infty} \lim_{N \to \infty}$ (anomalous thermodynamical metaequilibrium) and $\lim_{N \to\infty} \lim_{t \to \infty}$ (BG thermodynamical equilibrium) are not!
 interchangeable if $0 \le \alpha/d \le1$, whereas they are if $\alpha/d>1$.

Useful remarks from A. Rapisarda are gratefully acknowledged.
Partial support from CNPq, PRONEX, FAPERJ (Brazilian agencies),
and FCT (Portugal) is also acknowledged.

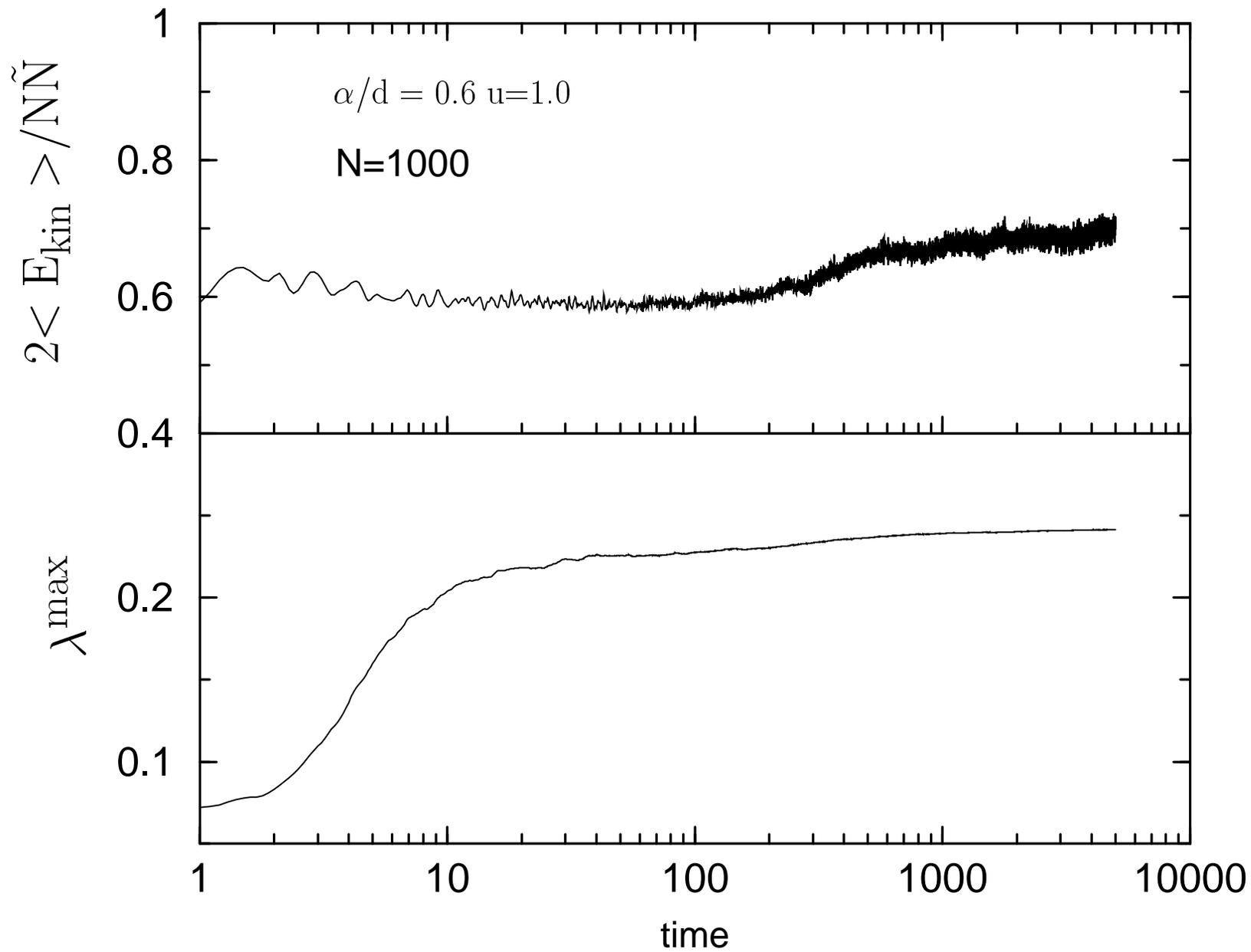
{\bf Fig. 1} - Time evolution for twice the (scaled) average kinetic energy
per particle $\langle E_{kin} \rangle /N {\tilde N}$
(which plays the role of temperature; upper curve)
and the (scaled) largest Lyapunov exponent
$\lambda_{u<u_c}^{max}$ (lower curve) for $\alpha = 0.6$,
$u=1$ and $N=1000$. We have averaged 10 different water-bag
initial conditions for the velocities
(all angles were initially set parallel to each other).

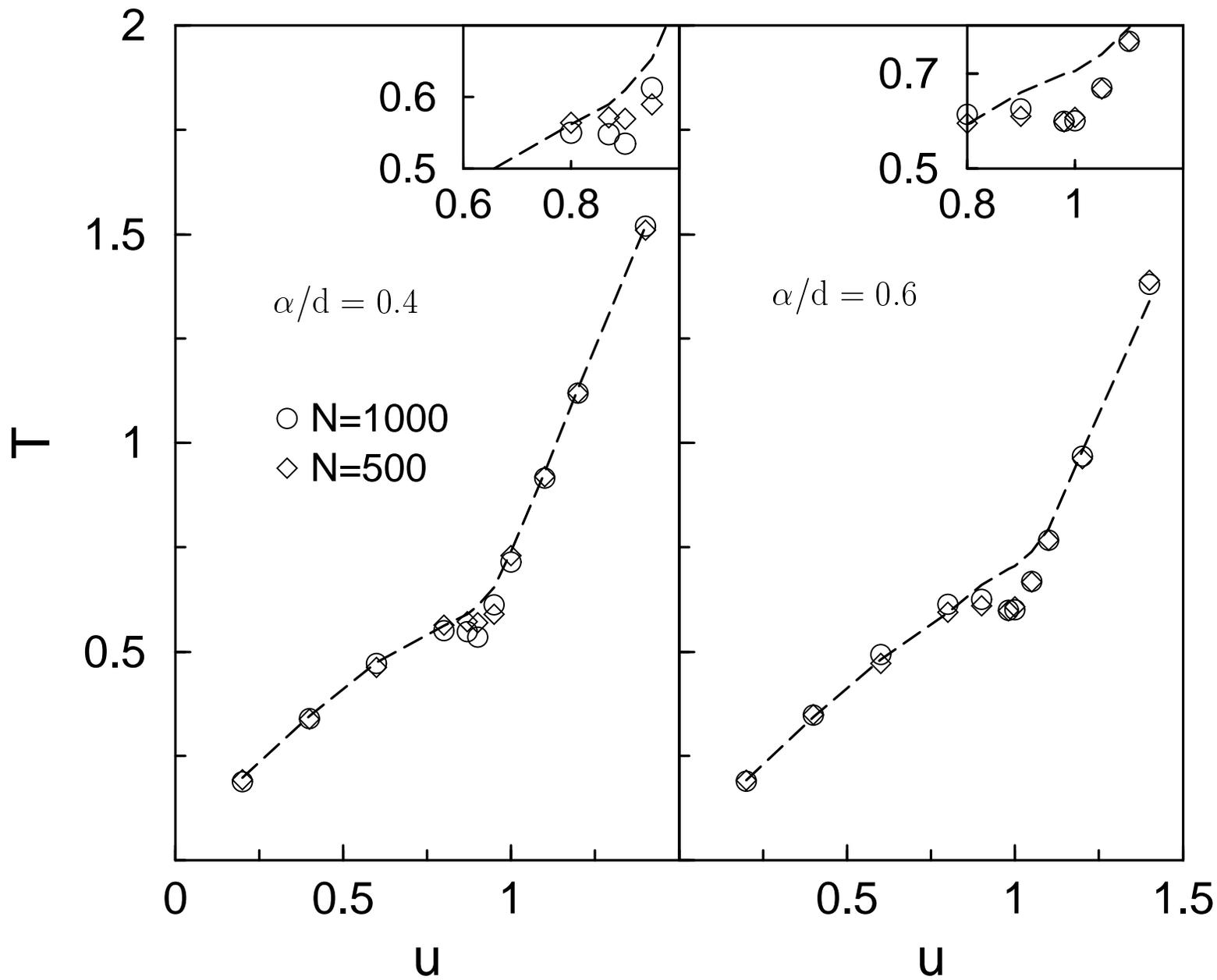
{\bf Fig. 2} - Microcanonical caloric curves for typical values
of $\alpha$ and $N$. The lower branch corresponds to the 
metastable state. The stable state is indicated with the dashed line.

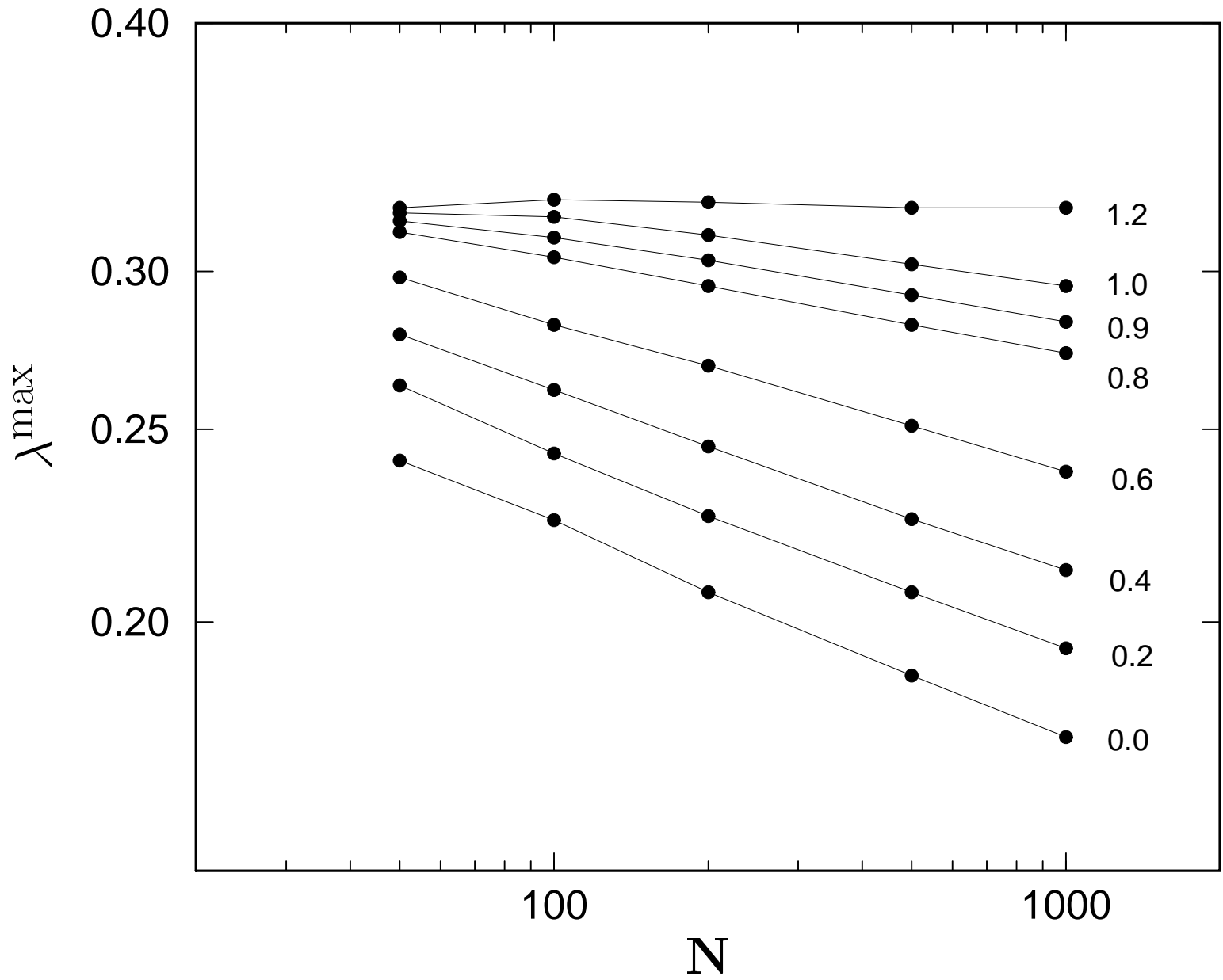
{\bf Fig. 3} - $N$-dependance of the largest Lyapunov exponent 
$\lambda_{u<u_c}^{max}(metastable)$ for $\alpha$ ranging from 0 to 1.2. The average in the interval $10<t<3000$ has been considered
as the metastable state value (the very slight increase of the Lyapunov exponent occasionally observed up to $t=3000$ is numerically  without consequences).

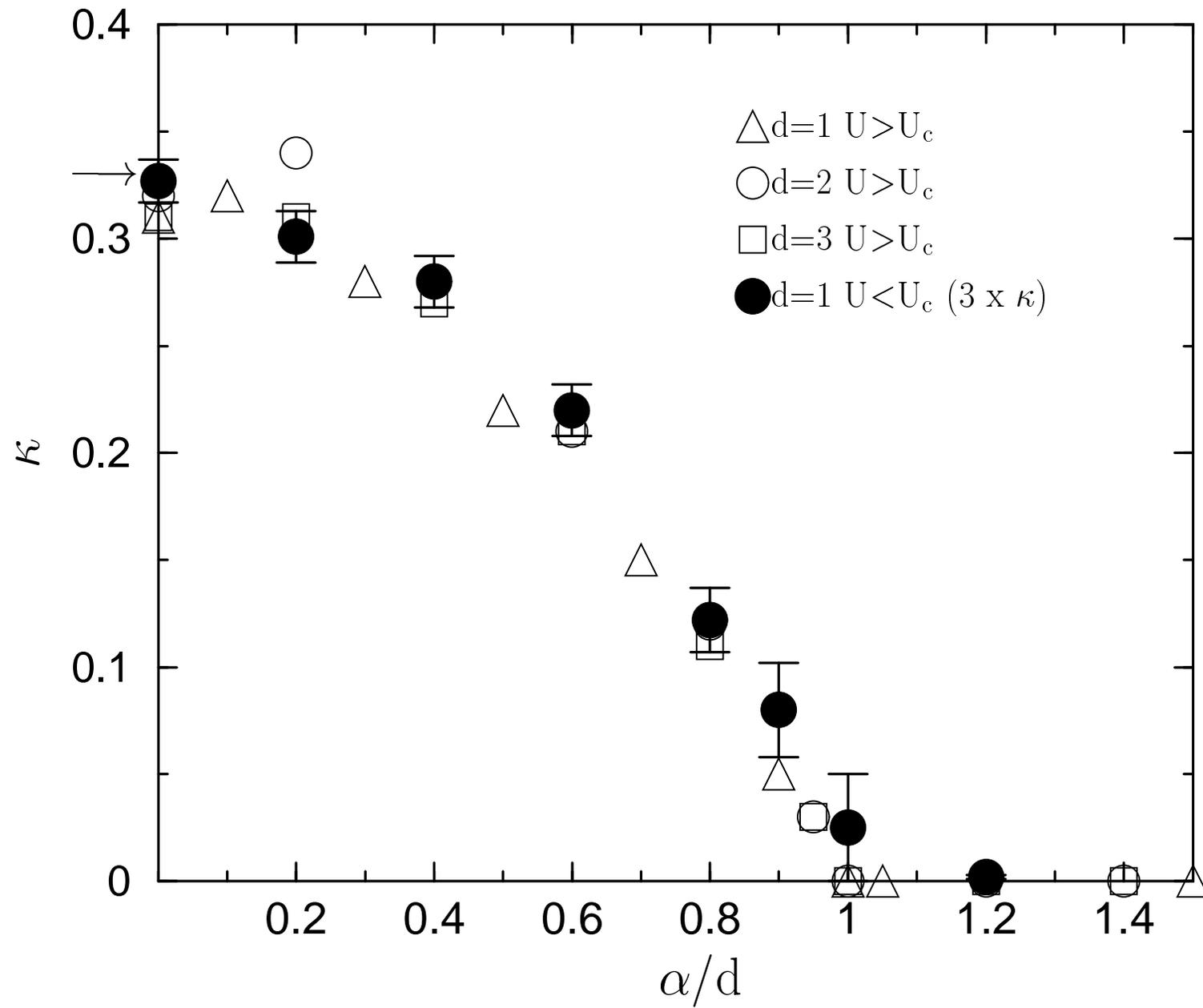
{\bf Fig. 4} - $\alpha/d$-dependance of $3 \times \kappa_{metastable}$
 (full cercles).
Open triangles, cercles and squares respectively correspond to 
$\kappa_{d}$ of the $d=1, 2, 3$ models \cite{anteneodo,giansanti}. The arrow
 points to 1/3, value analytically expected \cite{firporuffo,celiaraul} to
 be exact for $\alpha=0$ and $u>u_c$ .


\begin{thebibliography}{10}


\bibitem{einstein}A. Einstein, Annalen der Physik {\bf 33}, 1275 (1910) [
``Usually $W$ is put equal to the number of complexions... In order to
calculate $W$, one needs a {\it complete} (molecular-mechanical) theory of the
system under consideration. Therefore it is dubious whether the Boltzmann
principle has any meaning without a complete molecular-mechanical theory or
some other theory which describes the elementary processes.
$S=\frac{R}{\cal N}\log W+\;{\rm const}.$ seems
without content, from a phenomenological point of view, without giving in
addition such an {\it Elementartheorie}.'' (Translation: Abraham Pais, {\it
Subtle is the Lord...}, Oxford University Press, 1982)].


\bibitem{tsallis}C. Tsallis, J. Stat. Phys. {\bf 52}, 479 (1988); 
E.M.F. Curado and C. Tsallis, J. Phys. A {\bf 24}, L69 (1991) 
[Corrigenda: {\bf 24}, 3187 (1991) and {\bf 25}, 1019 (1992)]; C. 
Tsallis, R.S. Mendes and A.R. Plastino, Physica A {\bf 261}, 534 
(1998). For a regularly updated  bibliography of the subject see 
http://tsallis.cat.cbpf.br/biblio.htm . 


\bibitem{tsallisreview}S.R.A. Salinas and C. Tsallis (eds.), 
{\it Nonextensive Statistical 
Mechanics and Thermodynamics}, Braz. J. Phys. {\bf 29}, Number 1  (1999);
S. Abe and Y. Okamoto (eds.),  {\it Nonextensive Statistical 
Mechanics and Its Applications}, 
Series {\it Lecture Notes in Physics} (Springer-Verlag, Heidelberg, 2001);
P. Grigolini, C. Tsallis and B.J. West (eds.),
{\it Classical and Quantum Complexity and Nonextensive Thermodynamics},
Chaos, Solitons and Fractals {\bf 13}, Number 3 
(Pergamon-Elsevier, Amsterdam, 2002);
G. Kaniadakis, M. Lissia and A. Rapisarda (eds.),
{\it Non Extensive Thermodynamics and Physical Applications},
Physica A {\bf 305}, Number 1/2 (Elsevier, Amsterdam, 2002);
M. Gell-Mann and C. Tsallis (eds.),
{\it Interdisciplinary Applications of Ideas from
Nonextensive Statistical Mechanics and Thermodynamics}
(Oxford University Press, 2002), to appear.


\bibitem{turbulence}C. Beck, Physica A {\bf 277}, 115 (2000);
C. Beck, G.S. Lewis and H.L. Swinney, Phys. Rev. E {\bf 63}, 035303 (2001);
C. Beck, Phys. Rev. Lett. {\bf 87}, 180601 (2001);
T. Arimitsu and N. Arimitsu,  Phys. Rev. E {\bf 61}, 3237 (2000),
J. Phys. A {\bf 33}, L235 (2000) and Physica A {\bf 305}, 218 (2002).


\bibitem{bediaga}I. Bediaga, E.M.F. Curado and J. Miranda,
Physica A {\bf 286}, 156 (2000); C. Beck, Physica A {\bf 286}, 164 (2000).


\bibitem{cosmic}C. Tsallis, J.C. Anjos and E.P. Borges, astro-ph/0203258.


\bibitem{arpita}A. Upadhyaya, J.-P. Rieu, J.A. Glazier and Y. Sawada,
Physica A {\bf 293}, 549 (2001).


\bibitem{latorarapisardatsallis}V. Latora, A. Rapisarda and C. Tsallis,
Phys. Rev. E {\bf 64}, 056134 (2001).


\bibitem{HMF}V. Latora, A. Rapisarda and S. Ruffo, Phys. Rev. 
Lett. {\bf 80}, 692 (1998); Physica D {\bf 131}, 38 (1999); Phys. 
Rev. Lett. {\bf 83}, 2104 (1999); Physica A {\bf 280}, 81 (2000);
 see also M. Antoni and A. Torcini, Phys. Rev. E {\bf 
57}, R6233 (1998).


\bibitem{anteneodo}C. Anteneodo and C. Tsallis, Phys. Rev. Lett. {\bf 80}, 5313 (1998). 


\bibitem{fisher}M.E. Fisher, Arch. Rat. Mech. Anal. {\bf 17}, 377 
(1964); J. Chem. Phys. {\bf 42}, 3852 
(1965); J. Math. Phys. {\bf 6}, 1643 (1965); M.E. Fisher and D. Ruelle, 
J. Math. Phys. {\bf 7}, 260 
(1966); M.E. Fisher and J.L. Lebowitz, Commun. Math. Phys. {\bf 19}, 
251 (1970).


\bibitem{tamaritanteneodo}F.A. Tamarit and C. Anteneodo,
Phys. Rev. Lett. {\bf 84}, 208 (2000).


\bibitem{giansantiPRE}A. Campa, A. Giansanti and D. Moroni,
Phys. Rev. E {\bf 62}, 303 (2000).


\bibitem{giansanti}A. Campa, A. Giansanti, D. Moroni and C. Tsallis,
Phys. Lett. A {\bf 286}, 251 (2001).


\bibitem{firporuffo}M.-C. Firpo and S. Ruffo, J. Phys. A {\bf 34}, L511 (2001).


\bibitem{celiaraul}C. Anteneodo and R.O. Vallejos,
Phys. Rev. E  {\bf 65}, 016210 (2002).


\bibitem{giansanticagliari}A. Campa, A. Giansanti and D. Moroni,
Physica A {\bf 305}, 137 (2002).


\bibitem{thirring}W. Thirring, Z. Phys. {\bf 235}, 339 (1970).


\bibitem{lynden-bell}D. Lynden-Bell, Physica A {\bf 263}, 293 (1999). 


\bibitem{posch}Lj. Milanovic, H.A. Posch and W. Thirring, Phys. Rev. 
E {\bf 57}, 2763 (1998).


\bibitem{TCU}P. Klinko and B.N. Miller,
Physica A {\bf 305}, 258 (2002); B.N. Miller and J.L. Rouet,
Physica A {\bf 305}, 266 (2002).


\bibitem{sota}Y. Sota, O. Iguchi, M. Morikawa, T. Tatekawa
and K. Maeda, Phys. Rev. E {\bf 64}, 056133 (2001).


\bibitem{mapetc1}C. Tsallis, A.R. Plastino and W.-M. Zheng, Chaos, Solitons
and Fractals {\bf 8}, 885 (1997).
See also P. Grassberger and M. Scheunert, J. Stat. Phys. {\bf 26} 697 (1981),
T. Schneider, A. Politi and D. Wurtz, Z. Phys. B {\bf 66} 469 (1987),
G. Anania and A. Politi, Europhys. Lett. {\bf 7} 119 (1988) and
H. Hata, T. Horita and H. Mori, Progr. Theor. Phys. {\bf 82} 897 (1989).
 
\bibitem{mapetc2} U.M.S. Costa, M.L. Lyra, A.R. Plastino and C. Tsallis, Phys.
Rev. E {\bf 56}, 245 (1997). 


\bibitem{mapetc3}
M.L. Lyra, C. Tsallis, Phys. Rev. Lett. {\bf 80}, 53 (1998).


\bibitem{mapetc4}
U. Tirnakli, C. Tsallis and M.L. Lyra, Eur. Phys. J. B {\bf 10},
309 (1999).


\bibitem{mapetc5}
F.A.B.F. de Moura, U. Tirnakli and M.L. Lyra, Phys. Rev. E {\bf 62}, 6361
(2000).
 
\bibitem{mapetc6}
 V. Latora, M. Baranger, A. Rapisarda and C. Tsallis, Phys. Lett. A {\bf 273},
 97 (2000); U. Tirnakli, G.F.J. Ananos and C. Tsallis, Phys. Lett. A {\bf 289}, 
 51 (2001).
 See also J. Yang and P. Grigolini, Phys Lett. A {\bf 263}, 323 (1999).


\bibitem{mapetc7}
 V. Latora and M. Baranger, Phys. Rev. Lett. {\bf 82}, 520
 (1999).
\bibitem{mapetc8}E.P. Borges, C. Tsallis, G.F.J. Ananos and 
P.M.C. Oliveira, cond-mat/0203348.


\bibitem{mapetc9}F. Baldovin, C. Tsallis and B. Schulze, cond-mat/0203595.


\bibitem{tsalliscatania}C. Tsallis,
communication at the HMF Meeting (6-8 September 2000, Catania, Italy).


\bibitem{Yoshida} H. Yoshida, Phys. Lett. A {\bf 150}, 262 (1990).


\bibitem{latorarapisardacagliari}V. Latora, A. Rapisarda and C. Tsallis,
Physica A {\bf 305}, 129 (2002).


\end{thebibliography}
\end{document}